# Precision Medicine Informatics: Principles, Prospects, and Challenges

**Muhammad Afzal[1], Member, IEEE, S.M. Riazul Islam[2], Member, IEEE, Maqbool Hussain[1], and Sungyoung Lee[3]**

[1]Department of Software, Sejong University, Seoul, South Korea
[2]Department of Computer Science and Engineering, Sejong University, Seoul, South Korea
[3]Department of Computer Science and Engineering, Kyung Hee University, Yongin, South Korea

Corresponding author: Muhammad Afzal (e-mail: mafzal@sejong.ac.kr).

This research was supported by the Ministry of Science and ICT (MSIT), Korea, under the Information Technology Research Center (ITRC) support program (IITP-2017-0-01629) supervised by the Institute of Information & Communications Technology Planning & Evaluation (IITP). This work was supported by an IITP grant funded by the Korean government (MSIT) (no.2017-0-00655).

**Abstract** Precision Medicine (PM) is an emerging approach that appears with the impression of changing the existing paradigm of medical practice. Recent advances in technological innovations and genetics, and the growing availability of health data have set a new pace of the research and imposes a set of new requirements on different stakeholders. To date, some studies are available that discuss about different aspects of PM. Nevertheless, a holistic representation of those aspects deemed to confer the technological perspective, in relation to applications and challenges, is mostly ignored. In this context, this paper surveys advances in PM from informatics viewpoint and reviews the enabling tools and techniques in a categorized manner. In addition, the study discusses how other technological paradigms including big data, artificial intelligence, and internet of things can be exploited to advance the potentials of PM. Furthermore, the paper provides some guidelines for future research for seamless implementation and wide-scale deployment of PM based on identified open issues and associated challenges. To this end, the paper proposes an integrated holistic framework for PM motivating informatics researchers to design their relevant research works in an appropriate context.

INDEX TERMS: Precision Medicine; Bioinformatics; Informatics; Artificial Intelligence; Internet of Things; Big Data; Clinical Decision Support; Deep Learning; Machine Learning.

## I. INTRODUCTION

Precision Medicine (PM) is one of the fledging paradigms that the next generation healthcare solutions sprouting towards. It helps us grow more knowledge on human physiology by means of genomic insights and advances in technology. PM is an attention-grabbing area of research for medicinal community with various multidimensional prospects. At the same time, it is quite exciting for informatics community with enormous potential to research and exploit the technological perspective for the common goals. It is however challenging for either community to absorb the technicalities involved in drawing relationships among different prospects in this cross-disciplinary research field. From informatics viewpoint, PM introduces a new level of challenges on the developing informatics solutions including omic informatics and health informatics for a more focused and precise patient care.

### A. BRIEF OVERVIEW OF PRECISION MEDICINE

The concept of PM emerged as a healthcare-aligned mainstream discipline through its formal launching in 2015 as the prevention and treatment that consider the individual variability [1]. To put it simply, PM refers to serve the right patient with the right drug at the right time, by considering the molecular events that are accountable for the disease [2]. The term precision medicine is often muddled with personalized medicine [3], [4] due to the inclusion of the word "Individual" in the definition of the PM itself. However, the PM provides a more comprehensive and precise meaning to what individualized and personalized medicine were representing over the years. Unlike personalized medicine, the notion of PM is to combine clinical data with population-based molecular profiling, epidemiological data and other data so as to make clinical decisions for the benefit of individual patients [5]. The personalized medicine terms is used dominantly in some regions of the world and in a commentary, the authors termed PM as a part of personalized medicine [6]. The other terms they mentioned include "individualized medicine," "genomic medicine," "stratified medicine," "pharmacogenomics," and "P4 medicine". This study, however, uses the term "precision medicine" as a main subject in the search queries and focuses on the same in the contents to avoid any confusion with other competitive terminologies.

The paradigm shift to PM from the traditional medicine approaches can be thought of as a movement from generalization to personalization. In other words, unlike the current approaches that consider a general understanding based on the average conditions and integrated expertise on the different but interrelated domains including, to the minimum, physicians, biologists, and computer scientists. It is clear that two aspects of participation in PM are taken of utter importance: (i) the healthcare system in order to deliver precise diagnosis and therapies and (ii) the scientists to develop the infrastructure, principles, and insights into PM [9].

### B. STUDY OBJECTIVE AND CONTRIBUTIONS

In this study, we explore an informatics perspective of PM describing principles, issues, challenges and prospective solutions. Moreover, we include different initiatives around the world on the subject and a historical journey to clinical outcomes for the patients of interest, the PM approach works based on the individual variability in genes, environment, and lifestyle [4]. Consequently, whereas current approaches might be successful for one group of patients and not for the other, PM-based approaches are more likely to be effective for each group of patients. The abstract level comparison of PM with current approaches is depicted in Figure 1. The schematic shows the key differences between traditional and PM approaches in terms of classification of patient population – whereas PM classifies the patients based on risk and identifies the surveillance for preclinical disease, conventional approaches look for the signs or symptoms and deal the patients equally if they share the same symptoms [7]. Because of this generalization, in conventional approaches the benefits are not reached out to all the patients; however, in PM, each group of patients get equal level of benefits as they are treated rightfully with the right treatment.

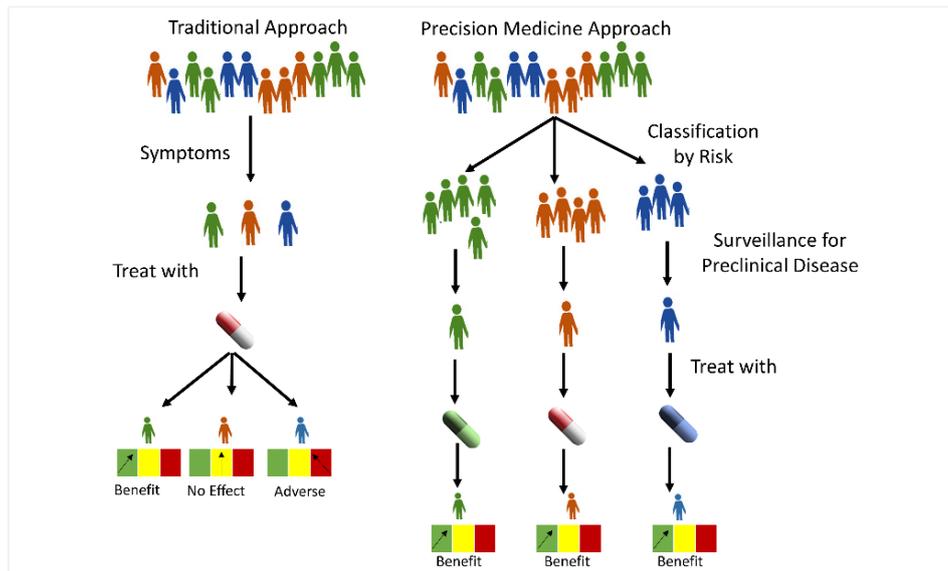

FIGURE 1. *Traditional and PM approaches with key differences on classification factors and treatment outcomes.*

The PM approach attracts multiple stakeholders in the biomedical enterprise, including care providers, payers, researchers, and patients [8]. Also, it seeks for the create a case for bridging the current evidence-based medicine (EBM) with PM. The existing studies [8][10] provide a big picture of an informatic research and envision the need of advanced tools and technologies to support PM. Also, we can find a fair set of literature [11][12] that discuss about the PM realization and implementation issues and challenges. The larger set of existing studies is available on the molecular and -omic information in terms of efficient algorithms and methods for genome mapping, alignments, variant callings, and annotations. Similarly, the clinical aspect has been researched and implemented in the long run without aligning the focus to consider the other aspects of PM - genome and environmental data. Moreover, PM is recognized as tantamount to a technology-driven approach

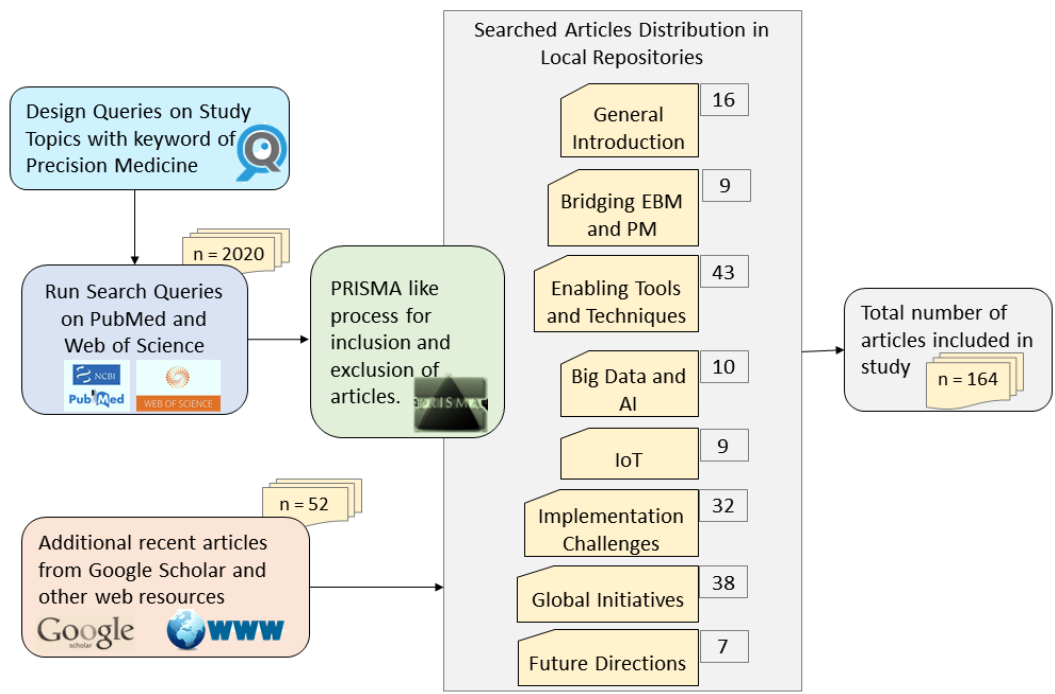

FIGURE 2. **PRIMA flow diagram for literature survey of the articles included in the study.**

[13], therefore, it embroils algorithm and technology in its meaning.

This study provides an overview of existing efforts on PM informatics agenda, tools and techniques in three areas of informatics – bioinformatic, clinical informatic, and participatory health informatic, security, standardization, integration, and implantation challenges, and the design of holistic PM framework to enlighten the futuristic endeavors in the area of informatic research and implementation. In this regard, the contributions of this paper are outlined below:

- To encourage the principle of 'learn to exist' rather than to compete, this study compiles the state-of-the-art views on PM to achieve a pragmatic balance among the existing approaches. The study adds on the reconciliation strategies between the existing evidence-based medicine (EBM) and emerging PM approaches.
- To cover an inclusive picture of PM from tools and technologies perspective, we elaborate and generate a comprehensive summary of prominent programs, tools, frameworks, and platforms in three aspects of informatics: bioinformatics, clinical informatics and participatory informatics.
- The lifelines of PM- Big data and artificial intelligence (AI) are included and elaborated in the study to draw a useful relationship model with PM.
- The internet of things (IoT)-enabled healthcare has potentials to be a part of PM. In this context, we briefly discussed advantages of IoT-aided PM and presented a conceptual model that integrates both the paradigms.
- The study analyzes the implementation challenges of PM and highlights the design issues of clinical decision support systems. It takes into account the integration and standardization challenges in terms of data privacy, safety, security, and exchange standards for interoperability, and issues of realization and design of an ecosystem for PM.
- Based on identified limitations on PM implementations, we propose a holistic integrated PM framework that assists computer scientists, health- and bio-informaticists to carry forward the challenge of successful realization of PM.

### C. LITERATURE SURVEY METHODOLOGY

Objectively, we employed the PRISMA (Preferred Reporting Items for Systematic Reviews and Meta Analyses) [14] method for literature survey based on the process followed in [15] with additional customizations in the inclusion/exclusion criteria. We ran search queries on two search engines viz. Web of Science and PubMed and linked all the search results into a local repository. A bulk of peer-reviewed articles are checked for duplication and the abstracts are screened to exclude all those articles that are focused either on biology, molecular and/or clinical perspective or unavailability of the full-text documents. The rest of the articles are checked for eligibility criteria to include articles focusing on the topics noted earlier with PM as a primary content. It should be noted that some of the articles are cited just for general reference on the topic even though the central content therein is not PM. For example, [16] that talks on IoT in healthcare is a topic-oriented citation rather than a PM-focused citation. Similarly, we also referred to few popular websites and blogs, particularly, where the contents were of introductory nature such as PM global initiatives. The Figure 2 explains the steps taken in the entire literature survey process. The number of articles excluded at different stages and the final set of articles included in the study are explicitly mentioned.

The rest of the paper is structured as follows. Section 2 describes the need of bridging the gap between EBM and PM. Section 3 explains the enabling tools and techniques of PM. Section 4 is dedicated to discuss the Big Data and AI in PM followed by section 5 that highlights the role of IoT in PM. Section 6 analyzes the implementation challenges while section 7 focuses on the global initiatives regarding PM. In section 8, we provide the future direction and presented our proposed integrated framework for PM. The final section concludes this systematic survey.

## II. BRIDGING EBM AND PM

EBM has long been utilized in healthcare environment to serve different purposes like supporting clinical decisions, medical education, and health awareness. According to the comprehended definition described in [17], EBM is the use of evidence collected from well-made research formulated in primary studies such as meta-analyses, systematic reviews, and randomized controlled trials used for improved decision-making in medicine. In this way, EBM approximating the "one size fits all" implies the scenario of applying to all although it may not be exact from the perspective of EBM proponents. As we learned in the preceding section, PM focuses mainly on the individualistic behavior, a deviating scenario from the EBM. As shown in Figure 3, there also exist differences between EBM and PM in terms of the basic elements in decision-making process. However, both share similar characteristics on multiple grounds. In an editorial, authors opinioned that EBM and PM can be more advantageous if they can adopt the principle of 'learn to exist' in a symbiotic relationship to attain a pragmatic balance between them [18].

They further hinted to an important factor of bridging the two paradigms – if we fail to do so it might turn out with non-integrable outputs to address the health requirements, they originally set out to address. Similarly, authors concluded in their study [19] that EBM and PM complement rather than oppose one another although these approaches have their own merits and shortcomings. However, the efforts to reconciling EBM and PM demand a clear understanding of the fundamental differences between them. We investigate the differences and similarities between EBM and PM and present the findings in Table I.

The co-existence of EBM with PM amid the differences mentioned in Table 1 raises several challenges in terms of volume, format, and structure of data. We turn out few of the challenges that are certainly required to be sorted out making the amalgamation of EBM and PM a success. In Table II, some of the challenges are presented with tentative solutions with the aim of bridging the two paradigms.

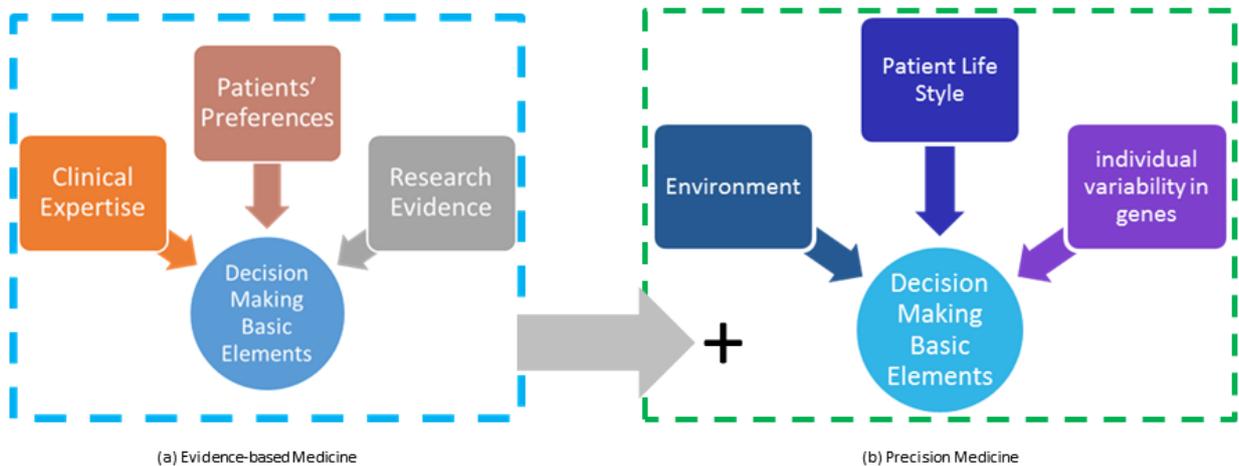

FIGURE 3. The decision-making basic elements of EBM in (a) and PM in (b); which indicates that all the elements in (a) are included in (b).

TABLE I
SIMILARITIES AND DIFFERENCES BETWEEN EBM AND PM

| Similarities |
|---|
| • Both EBM and PM have the objective of providing better decision on patient health problem. |
| • Both demand high quality and reliable evidence for the care of patients. However, the meaning of evidence could be different in two paradigms. |
| • Respecting patient in terms of either preferences or lifestyle is a part of both EBM and PM decision-making basic elements. |
| **Differences** |
| • EBM projects "one size fits all" approach and does not provide adequate solution for outliers. By contrast, PM deals with the outliers and projects the idea of "one size doesn't fit all" scenario [20]. |
| • EBM is cognitive-biased on occasions where clinicians set the goal and question for the trials and may favor the publication based on reputation, the product of manufacturer who funds the study to be conducted [21], [22]. PM, on the other hand, relies on patient information that are existed rather than to rely on hypothesis only. |
| • Since EBM relies on RCTs, outcome of RCTs are received in the form of either benefit, no effect, or adverse. In case of PM, the outcomes shall always be beneficial because they are target oriented that may leads to invent a new drug for the treatment [23]. |
| • EBM over-emphasizes the clinical consultation and is mainly concerned about the people who seek care. It underestimates the power of social networks where people can inform each other about their health problems [24]. Since it focuses on individual preferences, PM thus encourages the emerging ways of data curation from diverse sources. |

TABLE II
CHALLENGES AND POTENTIAL SOLUTIONS OF RECONCILING EBM AND PM

| Challenges | Prospect solution |
|---|---|
| • Analysis of voluminous data resided in different databases<br>• Bringing together data of various formats such as clinical and molecular.<br>• Lack of standardization of data entry and storage<br>• Understanding the paradigm shift from therapy to prevention, thus ultimately leading to clinician-to-patient communication and citizen-centered healthcare [25].<br>• Current published research has minimal patient input [24], thus it requires to include larger patient input in the future publishing. | • The proponent experts from both EBM and PM paradigms need to form a consortium/body to construct a unified architecture on the common grounds to revise the basic elements of clinical decision making.<br>• Revisions and update of the guidelines of developed for EBM, for instance, the criteria of RCTs structure, conducting, and evaluations.<br>• Devising a method to include patient input in the future research publishing. |

### III. PM Enabling Tools and Techniques

Precision Medicine introduces a new level of challenges for developing informatics solutions including –omic informatics and health informatics for a more focused and precise patient care. The informatics solutions range from data curation to processing, interpretation, integration, presentation, and visualization. The need for such enabling informatics solutions have been realized and discussed in array of studies [2], [26]–[30] with a central point of requirement of tools and techniques for voluminous, complex, and heterogeneous data processing, integration, and interpretation as well as knowledge acquisition and sharing. In order to describe the diverse set of PM enabling tools and techniques, we classify the tools in Figure 4 based on three areas of informatics: bioinformatics, clinical informatics, and participatory health informatics.

*A. Bioinformatics Tools*

Bioinformatics refers to the establishment of an infrastructure to provide means for storing, analyzing, integrating, and visualizing large amounts of biological data and related information and providing access to it using advanced computing, mathematics, and different technological platforms [31]. The term 'big data' resonates

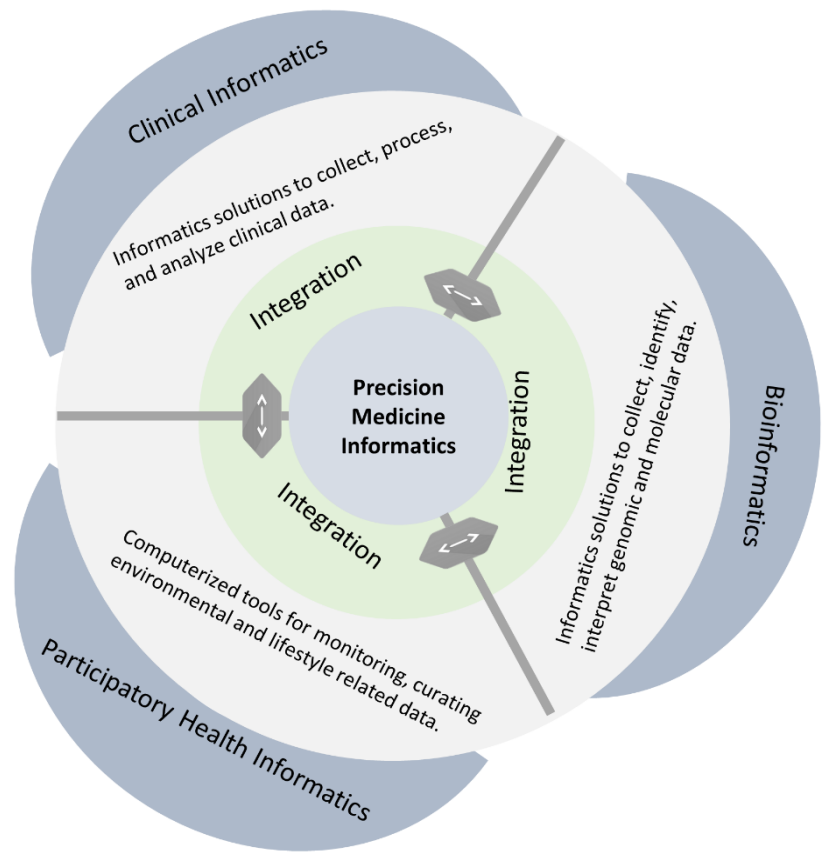

FIGURE 4. Classification of enabling tools and techniques in three areas of informatics: clinical informatics, biomedical informatics, and participatory health informatics

in the contents while talking about bioinformatics because big data technologies are required to process and analyze large genomic data sets [32]–[34]. Several tools are designed to work on genomic data at different levels. For instance, GMAP (genomic mapping and alignment program) [35] is a tool for next generation sequencing with a key functionality of sequence mapping and is designed to work on genomic, transcriptomic, and epigenomic data. Other similar tools used for –omic data processing, searching, and alignments include BWA (Burrows-Wheeler aligner) [36], STAR (spliced transcripts alignment to a reference) [37], GATK (genome analysis toolkit) [38], BLAST [39], DIAMOND [40] and others. The OmniBiomarker is a web-based application developed for biomarker identification that utilizes a curated knowledge base of cancer-genomic for analysis of high-throughput data [41]. Similarly, a number of tools are available for –omic data modeling such as, CODENSE (coherent dense subgraphs) [42], MEMo (mutual exclusivity modules in cancer) [43] and WGCNA (weighted correlation network analysis) [44]. The leading tools available for the molecular and biological data analysis among others are Geneious Prime [45], Cytoscape [46], [47], and Gephi [48]. There is a cloud-based integrative bioinformatics platform for precision medicine called G-Doc Plus for handling biomedical big data using its cloud computing resources and other computational tools [49]. On top of that, there exist a few other language-specific tools, libraries, and software packages in bioinformatics such as BioJava [50], BioPHP, BioPerl [51], PioRuby [52], and BioPython [53].

### B. Clinical Informatics Tools

Clinical informatics is a subfield of health informatics, which focuses on the usage of information in support of patient care [54]. Over the last two decades, clinical informatics has progressed with an array of tools and techniques including computerized entry systems, analytical tools, decision support tools, and other clinical reporting techniques appeared to assist healthcare professionals in different aspects. One of the related areas in clinical informatics is clinical information extraction, on which a significant volume of research had been conducted over the years. A methodological review [55] reported a wide range of information extraction frameworks, tools and toolkits including cTAKES [56], MetaMap [57], MedEx [58] and others. Another related aspect of clinical informatics is healthcare data analytics. The analytics area itself is huge and we do not necessarily aim to cover every aspect of it. However, the popular tools and techniques that focused on clinical aspects of the medical data are discussed. This is to emphasize that healthcare data analysis area is not limited to clinical data analysis rather it encompasses the combined analysis of phenotypes and genotypes. Data science tools in general have been used at the same pace and importance as of clinical data analysis. For instance, RapidMiner [59] and KNIME Analytics Platform [60] are the leading data science tools that are equally applied for clinical data analysis.

*C. Participatory Health Informatics Tools*

PM expands the scope of medical care as most of the population spends more time outside than the time in the physician's office. It demands a deeper consumer participation to collect information about a person's lifestyle and environment e.g. physical activity, dietary information, sleeping patterns, and other environmental conditions [8]. A significant number of systems and applications is added to the portfolio of quantified-self programs and digital health in recent years due to the increasing trends in wearables and mobile technologies [61]. Mining Minds (MM) project, is aimed at developing a novel framework for mining individual's daily life data produced from diverse resources [62]. The objective of the MM project is in line with other endeavors such as Google Fit [63], Samsung Health [64], Fitbit [65], and Noom [66]. However, it is more comprehensive and novel in terms of knowledge acquisition and context-aware personalized knowledge-based service support. Currently, these tools and services are geared towards nursing user health status for physical activities, diet, and somewhat sleep patterns and as well as the environmental factors. In the era of PM, it is required to channelize such efforts in a way to supply consumers' health monitoring and environmental information to their respective health providers for assisting in decision making. The genomic information could better be interpreted with this information and it will thus assist the physicians to precisely diagnose a disease and treat. There is a lot of other platforms and frameworks which cannot be put in a specific category rather then they are enablers for data analysis such as platforms for big data analytics including Apache Hadoop (MapReduce) [67] and IBM Infosphere Platform [68]. A selected list of tools for –omic data processing and biomarker identification is provided in a study on -Omic and EHR Big Data Analytics for Precision Medicine [29].

*D. Summary of PM Enabling Tools and Techniques*

The prominent tools and platforms to support PM in data processing, analysis, interpretations, sharing, and visualization reported in various studies are available under public and commercial licenses. It is important to note that some of the available platforms and tools are domain independent and are used for data analysis of any domain. For instance, the RapidMiner Studio [59] is a cross-platform data science tool, clinical informaticians use it for clinical data analysis and very recently, bioinformatics tools start integrating it in their workflows for enhanced data mining, analysis, and visualization.

## IV. Big Data and Artificial Intelligence

In medicine, the application of AI has two divisions: virtual and physical. Whereas the former is characterized by machine learning (ML) and/or deep learning (DL), the latter comprises physical objects, medical devices, and sophisticated robots [69]. In fact, the use of big data and AI is a central aspect of PM initiatives [70] and some even phrased it "there is no PM without AI" because of its fundamental requirement of computing power, algorithms machine learning and deep learning), and intelligent approaches that uses the cognitive capabilities of physicians on a new scale [71]. Deep learning has been widely used for clinical information extraction, phenotype discovery, image analysis, and next generation sequencing [72], [73]. AI upsurges learning abilities and offers decision-making capabilities at a scale to transform the healthcare future [74]. Therefore, physicians in everyday practice get pressure to look around innovations spreading over faster than ever through the use of disruptive technologies and exponential growth of healthcare data- the "Big data" [75]. Big data has gained a growing attention from data-oriented enterprises in private and governmental sectors [76]. Despite the fact that we are living the in the age of big data, however, the big data by itself is of no use without the processing using AI techniques which make it useful thus brings the potential to transform the current clinical practice [77]. AI techniques, such as applications of machine learning on big data, are changing the way physicians make clinical decisions and diagnosis. Big data analytics using PM platforms has therefore the potential to include data of millions of patients for exploration and validation [77]. It is of more importance to understand interrelationships among big data, AI (ML and DL), and the PM. Developing a PM platform or relevant tools and services requires access to big data and processing (big data needs AI approaches including ML and DL variants. Figure 5 illustrates the relationship of big data and AI in PM derived from the illustrations presented in [70], [77].

Many technology companies including IBM with a flagship platform of IBM Watson, Google with DeepMind, and others such as Apple and Amazon, are investing heavily in health care analytics to facilitate PM [70], [71], [77]. Despite the facilitation and improvements powered by AI for genomic and other omic data processing and analyzing, there still exist various challenges. In a review [78], authors focus on AI applications of next generation sequencing and cancer genomics testing required for PM. In Table III, we gathered a set of key benefits of AI in the era of PM and the associated challenges.

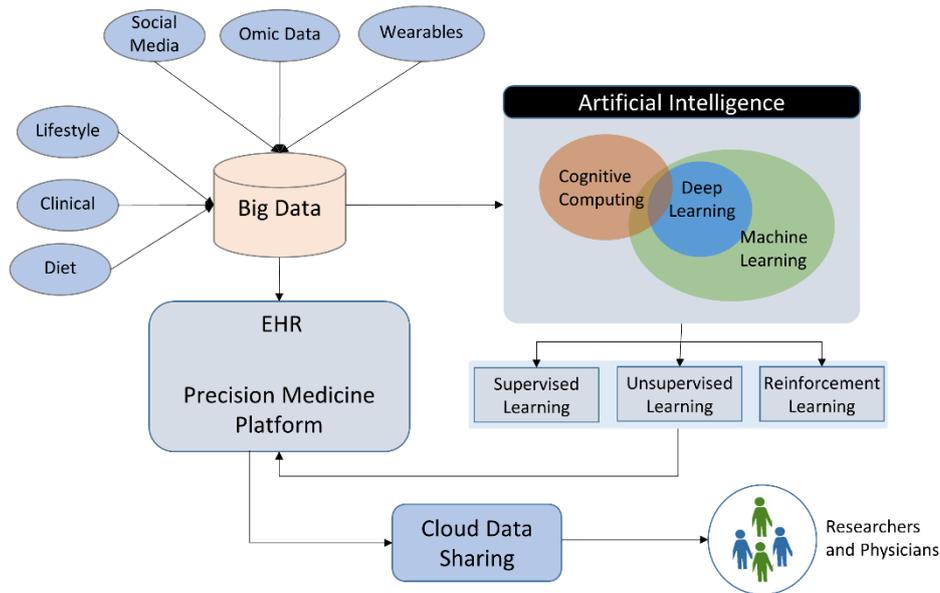

FIGURE 5. **Illustration of relationship between big data, artificial intelligence, and precision medicine**

TABLE III
KEY TOPICS WITH BENEFITS ON THE LEFT (GREEN ICONS) AND CHALLENGES ON THE RIGHT (RED ICONS)

| Key benefits of AI in PM | Challenges |
|---|---|
| **Variant calling** — AI algorithms leveraged to enable variant calling from NGS data. **Literature Mining** — AI algorithms are utilized for entity and relation extraction from published literature. **Variant interpretation and reporting** — AI algorithms are used to facilitate the process of variant classification and to ease manual curation. | **Ground truth scarcity for validation of benefit** — Obtaining statistically significant patient outcome data is challenging. **Transparency and reproducibility** — Companies, platforms, and publications offer limited information for public consumption. **Patient / physician education** — Both patient and physicians should get precision medicine related education to enjoy the outcomes brought by AI and big data. |

## V. Role of IoT in Precision Medicine

The internet of things (IoT) enables us to introduce automation in nearly every field and healthcare is one of the most important and attractive application areas of this auspicious technology. The IoT uprising is reshaping modern healthcare with propitious technological, economic, and social prospects. The role PM plays can be further enhanced by integrating the IoT.

As noted in the early part of this paper, PM primarily involves three categories of data – clinical data, genome data, and environmental data. On the other hand, a simple and brief description of how an IoT-based healthcare system works can be presented as follows. First, the IoT medical sensors and devices directly connected to the patient's body of interest. Sensors collect various physiological conditions and vitals. Accumulated data are preprocessed and organized and are subsequently analyzed. The data are stored in the associated medical service provider's cloud storages for aggregation. Depending upon the analytics and aggregation results, patients can be monitored from distant places and necessary actions are taken following predefined standard rules and guidelines. Interested readers are referred to the compressive study reported in [123] to grow more knowledge on the IoT-based healthcare. Clearly, the IoT can prominently assist PM by arranging the environmental data in an automated fashion because the participating health data is mostly collected by physical sensors and actuators. In addition, a dedicated intelligent coordinator can exploit the cross-sectional data consisting of IoT-provided data and clinical/genome data. Here we present an overview of several possible avenues of integrating the IoT with PM.

### A. Risk Minimization in ADR
The IoT can play a significant role in mitigating the risk associated with adverse drug reaction (ADR) [123]. In layman's terms, an ADR is an undesirable or injurious response experienced following the administration of a drug or combination of drugs under natural conditions of use and the suspicion of the unwanted response is held accountable mostly to the drug/s administrated [40]. A substantial number of admissions to hospital are caused by ADRs and hospitalized patients often experience ADRs that muddle and extend their stay. Many of the ADRs can be avoided if the appropriate care is taken. An ADR will usually require the drug to be discontinued or the dose reduced. For example, a simple sensing system can detect whether the dose or plasma concentration has risen above the therapeutic range. Such a concept of an IoT-based ADR is found in [74]. The work makes use of barcode or NFC-enabled devices so that the patient side recognizes the drugs. Then, an AI-based pharmaceutical system senses and analyzes the patient's health and molecular profiles. Eventually, the system performances matching comparison to conclude whether the suggested drug is well-suited. In a normal clinical viewpoint, the nature of ADR is characteristically generic i.e., not medication-specific for a particular disease. Therefore, a generic software package termed ADR services is required. The ADR services is supposed to cover certain mutual technical issues and their generic solutions [123]. However, the ADR services to be used in PM should be further customized and fine-tuned to cover the respective PM cohort.

### B. Safe and Secure Medication
The safety of the medicine in PM is one of the unique challenges that must be addressed by the pharmacists [79]. Also, the need for an entirely connected and transparent global healthcare supply chain will continue to grow and this is where the IoT can be useful. The IoT devices can monitor a bunch of parameters, including location, temperature, light exposure, humidity, as well as security to guard against theft and forging. Although this sort of supply chain control and monitoring is important for all industries, it is more vital for healthcare industry in general and PM in particular. For example, it will not be possible to have a quick substitute when a shipment of medicine that is personalized for the DNA of a patient with a life-threatening illness is spoiled or stolen.

### C. Medical Error Minimization
The main objective of PM is to delivery of optimized targeted stimulation. This is optimized, in the sense that the therapy is tailored to individual patients. The targeted stimulation does not allow a medication (e.g. taking a pill) to be metabolized throughout the patient's body. Instead, it stimulates the intended target in a controlled manner, and thereby reducing any side-effects. With the use of medical IoT devices, it is possible to steer the stimulation to a particular target with a much higher degree of precision [80]. As experienced in any system, the occurrence of medical error in healthcare in general is also affected by a several factors. With the introduction of PM, this error margin increases exponentially because of modular clinical treatment approaches. For example, caregivers (e.g. hospitals) are usually at over-capacity and thus they face scalability issue to increase access to care. Co-morbidity supervision becomes even more difficult than before. To address this issue, we can establish an IoT-based health network for an automatic patient caring process [81].

### D. Automation in Gene Expression Measurement
Gene expression profile has widely been used to uncover the association of environmentally-swayed or disease phenotypes with the mRNA expression patterns [82]. Due to its incredible application in computable genotyping, genetic variation of inter and intra organisms, early finding of disease, polymerase chain reaction (PCR) [83] and its subsequent derivatives are widely used to obtain real-time gene expression profiling. Then, because of rapid progress in miniaturized electrochemical DNA biosensors, it is possible to generate transformed electronic signal from the sensitive bio-receptor through a transducer (e.g. photo counter) in an automatic process, calling the need of an IoT-based health network, with a minimum involvement of technical personals in the close loop system. The system as a whole can eventually assist PM to predict disease risks and even what foods to consume based on patients' genome and extracted physiological sensors data [84].

The PM basically provides customized healthcare solutions to the individual cohort of patients. With the help of IoT, this customization itself can be improved by learning the individual's concerned physiological functions. For example, one can consider a possible way to improve the symptoms of a Parkinson's disease patient through a better deep brain stimulation (DBS) therapy using IoT (Figure 6).

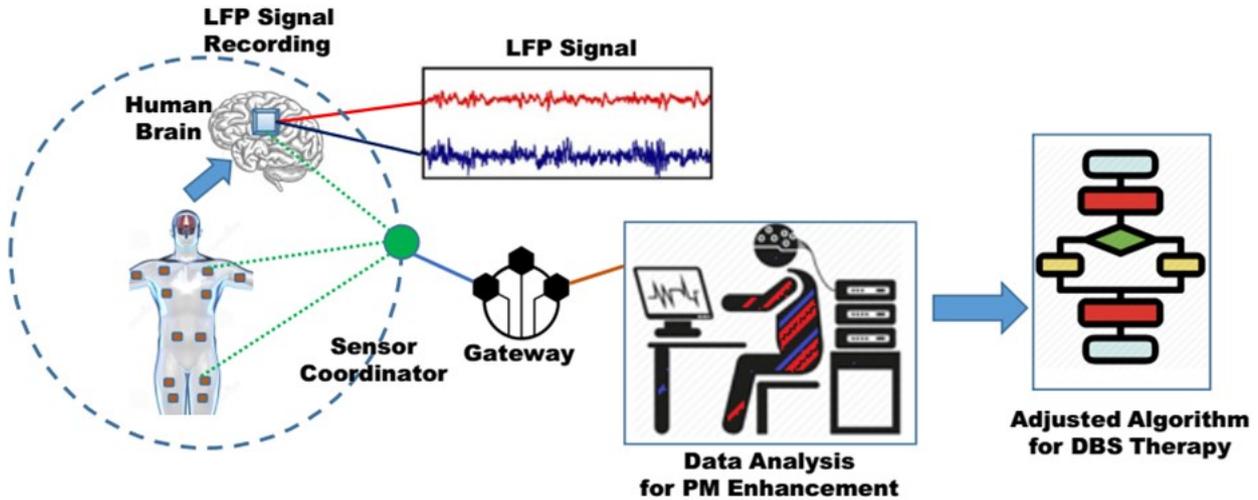

FIGURE 6. *A conceptual model of an IoT-aided personalized DBS therapy.*

DBS is a neurosurgical procedure which uses a neurostimulator that delivers electrical stimulation, through implanted electrodes, to specific targets in the brain for the treatment of neuropsychiatric disorders [85]. To advance the DBS therapy, we need to understand how individual's brain works. By sensing signals from the brain, we can learn more about how brain responds to the therapy. Low field potential (LFP) recording is promising to enable detection, measurement, and collection of brain signals [86]. The collected LFP signals along with other medical sensors data can then constantly be analyzed to improve the targeted DBS therapy [80]. This technology would eventually enable a precise adjustable algorithm which could lead a better understanding about various overwhelming neurological problems.

### E. Summary and Insight

Digital innovations collectively appear as a paradigm changer of how healthcare organizations provide quality patient services with enhanced clinical satisfaction while maintaining a safe and secure environment at each stage associated with the ecosystem. In line with that, IoT can potentially be integrated with PM to achieve improved automation in general. In particular, the IoT-enabled PM is potential to offer several benefits such as real-time monitoring of adverse drug reactions and secured healthcare supply chain. Even, IoT can be utilized to design innovative personalized therapies e.g., IoT-aided personalized DBS therapy, conceptualized in the early part of this section.

## VI. Implementation Challenges

In this section, we discuss more of the design challenges of futuristic PM system and services such as clinical decision support systems, ecosystem, and the challenges exist in in the integration and standardization of data elements and processes.

### A. Redesign of Clinical Design Support

Computer-based clinical decision support (CDS) are meant to enhance the decision-making capabilities by utilizing the individual-specific information and clinical knowledge [87]. It serves to facilitate different stakeholder like physicians, nurse, patients, and others in making effective clinical decisions. Formally, a CDS is referred to "a process for enhancing health-related decisions and actions with pertinent, organized clinical knowledge and patient information to improve health and healthcare delivery" [88]. Historically, the CDSs delivered promising results in diverse systems and services such as the reminder systems, the drug dosing and drug-drug interactions, the diagnoses and treatment, and the pharma-related fields [89]–[91]. Despite its potentials in improving health and healthcare, CDS has several challenges to accomplish its full promise [92]. Moreover, the fresh developments in the medicine domain and the presence of disruptive technologies pose a new set of challenges to develop models for CDS. This led us to put question to ask, do we need to rethink about the CDS's design in order to build a practical model for the PM era? The answer is certainly positive based on the realization by the researchers in their research works [8], [91], [93], where they pointed out the need of a CDS design that encompass a more comprehensive knowledge base (KB) to fulfill the key requirements of PM. Contemporary CDSs serve a fraction of clinical care whereas a common decision in PM shall require accumulating data from different components that are not integrated at one place. Researchers working in the area of informatics to advance PM [8] stressed upon the designing of an all-inclusive KB comprising information about disease subtypes and risks, diagnosis, treatment, and prognosis. Nevertheless, the available KBs are isolated from each other and are thus unable to provide support for executing the federated queries. On that, in addition to flexibility and scalability, the KBs need to be revamped to support not only the federated queries but also an extended reasoning capability. In a study [91], authors pointed-out the data isolation issue by highlighting the fact that the two sets of data, clinical and scientific, are typically placed in different repositories as information silos. They need to be linked and presented in a way that clinicians and other researchers can easily interact and review. This raises the requirement for a standard language and algorithm for executing a federated query. The Clinical Pharmacogenetics Implementation Consortium (CPIC) Informatics Working Group is developing a standard guidelines for the effective implementation of Pharmacogenetics in the day-to-day medical care [94]. This group also uncovered the limitation of present-day CDS issue of addressing single gene by relying on local versions of national guidelines. The group emphasized to step-forward to a national implementation by designing and implementing futuristic resources.

We summarize the core limitations of the contemporary CDSs in Table IV. Addressing these limitations while redesigning the CDSs in the era of PM, raises to a few challenges to consider for their resolution. In a study [89], we envisioned the conceptual architecture of the futuristic CDS eligible to support the functional requirements of PM services. In contrast to contemporary architectures of CDSs, the futuristic CDS model incorporates the modules of supporting federated queries, a supervisor KB that holds the information of disease subtypes and risks, diagnosis, treatment, and prognosis.

TABLE IV
LIMITATIONS OF CURRENT CDSs WITH CHALLENGES AND PROSPECT SOLUTIONS.

| Current CDSs Limitations | Challenges | Prospect Solutions |
|---|---|---|
| The knowledge bases of contemporary CDSs are isolated from one another and they are unable to support federated querying [8]. | How to design a comprehensive KBs to integrate information about various features such as disease subtypes, disease risk, diagnosis, therapy, and prognosis. | • Allowing data sharing and consensus on clinical interpretations and multiscale data.<br>• Enabling effective ontological modeling, knowledge provenance, and maintaining the integrated KB.<br>• Utilizing a set of novel computational reasoning approaches to allow efficient federated queries. |
| Isolation between scientific and clinical data [91]. | How to establish a meaningful connection between patient data and primary literature when the EHR databases are considered as information silo themselves? | • Applying standard vocabularies and data formats to integrate disparate data sources.<br>• Developing new research platform with a set of methods and tools to enable analysis and visualization of not only a massive amount of raw data generated in clinical set ups, but also the data resides in different databases of biomedical literature. |

## B. Design of PM Ecosystem

As noted above, we briefly mentioned about the three core aspects of informatics (i.e., bioinformatics, participatory health informatics, and clinical informatics) with their basics and provided information on selected set of tools and techniques. Nevertheless, it is also important to discuss the informatics solutions for PM which requires a holistic overview of working together as outlined below:

- Curation of data generated via participatory health using mobile devices, sensors, social media, and other IoT devices as well as environmental factors' data at a point of care for the assistance of genomic data interpretation which ultimately could help in precise patient care.
- Creating a synergy between bioinformatics and clinical informatics by developing infrastructure, tools, techniques and applications that bridge the two areas and allowing the sharing of data to offer integration of individual patient data into the clinical research environment [95].
- Development of a comprehensive framework that facilitates tools and techniques to integrate, process, and analyze data curated from diverse sources in all three areas; clinical, genomic, and lifestyle & environmental factors to enable one-point decision in a precise manner.
- Development of a coherent framework for dealing with multi-scale population data including the phenome, the genome, the exposome, and their interconnections [96].

A series of efforts have been made to provide informatics solution to support PM in a comprehensive manner. For instance, the network ENIGMA (Evidence-based network for the interpretation of germline mutant alleles) [97] is an international consortium for assessing clinical significance and risk related to sequence variation in genes, BRCA1 and BRCA2, which currently include over 100 research scientists and clinicians from 19 different countries. Similarly, ClinVar [98], an archive partner of ClinGen project is an archive (freely available) for variants' clinical significance interpretations. National Center for Biotechnology Information (NCBI) has provided an explorer tool to facilitate identification of clinical significance discrepancies in ClinVar [99]. In one of the reports [11] there is given a set of fundamental aspects of PM and describes the key aspects of computational infrastructure built on clinical-grade genomic sequencing. Authors therein emphasized on the integration of PM program into a medical institution's clinical system to facilitate billing and reimbursement. The proposed PM infrastructure integration with existing electronic health record infrastructure is shown in Figure 7. The existing EHR infrastructure depicted on the left is integrated with PM infrastructure on the right through passing the patient specimen information to the laboratory information management system (LIMS) in order to process, sequence, and analyze the specimen data. The LIMS component of the PM infrastructure sends back the report formed over the specimen data to the pathology system of the EHR infrastructure.

## C. Integration and Standardization

For successful data integration and exchange, data and metadata standards are required. However, there are several issues to achieve this goal in terms of either lacking of such standards or inconsistent use of existing standards, particularly in "omics" domain [8]. Prior to frame these issues for discussion, we first describe the meanings of what constitutes a 'data standard' in order to avoid confusion as different groups and individuals have different definitions for standards. According to the International Organization for Standardization, a standard is, '… a document that provides requirements, specifications, guidelines or characteristics that can be used consistently to ensure that materials, products, processes and services are fit for their purpose' [99]. There is further division in standards as data standards for integration and exchange and data standards for security, privacy, and integrity, covered as one of seven key areas in the research work [8]. The research emphasized on the need of extending the scope of the existing standards rather to invent a new. For that, the individuals or organization seek to adopt an existing standard should work closely with the owner of that standards to extend the scope. Not only co-working, relevant stakeholders should focus on outreach and education/training to educate the potential adopters of understanding and using existing data standards. Data standardization for integration and exchange is required for correct interpretation of the data elements. The motivation for data standardization on security and privacy comes from the notion of developing mutual consensus on the level of data as well as the protocol definition for sharing.

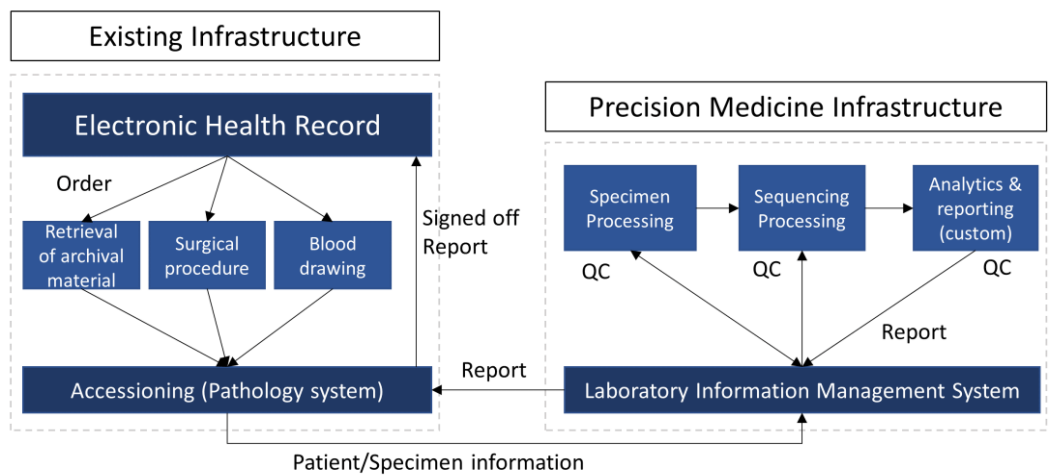

FIGURE 7. **Components of an integrated precision medicine workflow illustrating existing HER and novel PM infrastructure- (QC: Quality Control)**

A number of initiatives have been taken place to facilitate the adoption of data standards especially in the 'omics' discipline. One of these initiatives is BioSharing that works to ensure the standards are searchable and informative by mapping the landscape of community developed standards in the life sciences including biomedical sciences [100]. BioSharing facilitates those who are looking for information based on the existing standards, finding duplications or gaps, encourage harmonization to avoid reinvention, and developing criteria items for evaluating standards for adoption. The American College of Medical Genetics and Genomics (ACMG) together with Association of Molecular Pathology (AMP) and College of American Pathologists members formed a workgroup with the goal of developing a classification of sequence variants using criteria informed by expert opinion and empirical data [101]. This workgroup is aimed at providing detailed variant classification guidance to update the recommendations on interpretations of sequence variants previously provided by the ACMG. The Canadian Open Genetics Repository is an endeavor aims to establish collaboration of Canadian laboratories with other countries to support the development of tools for sharing laboratory data in addition to the collection, storage, sharing and robust analysis of variants in the laboratories across Canada [102]. There are other initiatives have been established to ensure data security and privacy through standardization. For instance, a framework is established that provides guidance for responsible sharing of genomic and health-related data including personal data [103]. The Minimum Information About a Microarray Experiment (MIAME) [104] provides guidelines for the minimum information to describe the experiment details of DNA microarray data so that the experiment could either be reproduced or analyzed the data de novo [105]. Similarly, data modeling and XML-based exchange standards 'Microarray Gene Expression' appeared in the form of Object Model (MAGE-OM), Markup Language (MAGE-ML) [106], and a Controlled Vocabulary called the MGED Ontology [107]. These standards resulted the creation and evolution of a several interoperable databases and repositories [105].

For seamless integration and information exchange, recently Health Level Seven (HL7) extended the efforts on Genomic data working group with its newly invented popular standard called Fast Health Interoperability Resource (FHIR) [108], [109]. One of the main FHIR resources is a Sequence resource which is designed to describe an atomic sequence containing the alignment sequencing test result and multiple variations [110]. For the facilitation of standardized clinic-genomics apps, a framework called Substitutable Medical Applications & Reusable Technologies (SMART) on FHIR Genomics is developed which specifies genomic variant data resource definitions [111]. SMART on FHIR Genomics specification offers developers a unified framework to work with multiple resources of genomic and clinical data to facilitate the type of apps required for precision medicine. A brief summary of pertinent initiatives on standardization in the area of PM is provided in Table V.

TABLE V
CLASSIFICATION OF STANDARD INITIATIVES IN THE DOMAIN OF PRECISION MEDICINE

| Initiative Name | Year | Standard Body | Scope |
|---|---|---|---|
| Biosharing [100] | 2011 | FAIRsharing team, Oxford e-Research Center. | Integration and Exchange |
| MIAME [104] | 2001 | FGED-Functional Genomics Data Society | Integration and Exchange |
| SMART on FHIR Genomics [109] | 2015 | HL7 ® International | Integration and Exchange |
| PMI Data Security Principles Guide [112][113] | - | ONC – Office of the National Coordinator for Health Information Technology | Privacy and Security |
| PMI-AURP [114] | 2016 | GSWG | Privacy and Security |
| HGNC database [115] | 2007 | HUGO | Vocabulary and/or Nomenclature |
| UMLS ® [116] | - | U.S. National Library of Medicine (NIH) | Vocabulary and/or Nomenclature |

### D. Summary and Insight

At a granular detailed level, there exist a lot of implementation challenges that are highlighted in various studies ranging from genotype data preprocessing, mapping and alignments, unstructured clinical text processing, image processing and environmental data acquisition and synchronization. In this section, we focused on generalized implementation challenges that exist irrespective of individuals' realizations of PM. The challenges are mostly related to the rethinking on a new design for the clinical decision support systems to include information from the other aspects of PM – molecular, -omic and environmental in order to produce the right decision for the right patient. We included the designs in the existing studies and provided the abstract representation of an ecosystem of PM for enhancing the design of existing electronic health record systems with genome. We deliberated the integration challenges at data and process levels and the standardization efforts at global spectrum.

### VII. PM Global Initiatives

PM is spreading globally with a fast pace and is thereby creating a multibillion market. According to a report issued by Persistence Market Research, global PM market is expected to approach $ 172.95 Billion by the end of 2024 [117], [118]. Different countries share this market by initiating innovative projects to support PM in terms of establishing infrastructure, research centers, working groups, and standardization bodies. Different countries allocated different sort and amount of funding to support the PM initiative. In this section, we briefly elaborate country-wise initiatives with their goals and way of working.

#### A. United States

United States of America took the lead launching the idea of Precision Medicine under the Obama administration back in 2015 [1]. The idea was taken further by the National Institutes of Health (NIH) and other partners. Initially, a budget of $215 million investment has been announced in the President's 2016 Budget. NIH-funded resource ClinGen is dedicated to constructing an authoritative central resource to establish clinical relevance of genes and variants for the convenient use in precision medicine and research [119][120]. Ensuring the accuracy of NGS tests, US FDA is working on three aspects; guidance for Databases to allow developers to use data from FDA-based databases of genetic variants, recommendations for designing, developing, and validating NGS tests, and support to develop bioinformatics tool to engage users across the world to experiment, share data, and test new approaches [121]. PrecisionFDA [122] as a community platform for NGS assay evaluation and regulatory science exploration, has resulted as an outcome of FDA efforts.

#### B. China

Precision Medicine is included as part of China's five-year plan with an expected investment of more than $9 billion for research. Among 40 countries where there are initiatives related to PM, China is on the top from investment perspective. Compared to the United States PMI investments, China is spending $43 for every $1 of US, thus making China as a global leader in PM [123]. The Beijing Genome Institute (BGI) [124] is the world's largest genomic organization with a focus of genetic sequencing. Affiliated to BGI, the China National GeneBank [125] has over 500 million genetic sequences stored in more than 40 databases, as of early 2017. Sichuan University's West China Hospital which is ranked the first among all Chinese hospitals for four consecutive years in science & technology influence, plans to sequence 1 million human genomes itself [126], [127]. AliCloud by Alibaba Group partnered with BGI and Intel Corporations have launched Asia's first cloud platform for precision medicine and its applications with a vision to accelerate the advent of precision medicine [128]. More to PM initiative, it is anticipated that leading institutes, including Fudan University, Tsinghua University, and the Chinese Academy of Medical Sciences, are trying to establish precision-medicine centres [126].

#### C. United Kingdom

United Kingdom has started a well-known project '100,000 Genomes Project' in 2013 with a goal to sequence 100,000 genomes from around 70, 000 people from the participants of National Health Service (NHS) patients having a rare disease [129][130]. This project is considered currently as the world's largest national sequencing project of its kind. A program Coordination Group led by Innovative UK is active in precision medicine to bring together representatives from UK public sector and charity funders and through this group, a dataset of over 400 infrastructure investments in precision medicine has already been developed [131], [132]. Innovate UK is envisioned to invest up to £6 million in precision medicine technologies related innovation projects [133]. Overall, the UK government has invested more than £1 billion in developing precision medicine research infrastructure [134].

#### D. Japan

Like other countries, Japan is also contributing to support the development of personalized and precision medicine (PPM) [135]. Japan has established three biobanks to collect genome data; Bio Bank Japan, National Center Bio Bank Network, and Tohoku Medical Megabank. All these three banks work together, however Bio Bank Japan being the largest of the thee, plans to collect data from 300,000 people alone. The total budget of $ 103 million is allocated in 2016 for the plans like clinical trials, research on genomic care, and establishing seven core hospitals to support the provision of genomic medical treatments [135]. Additionally, Japan has established the most successful National Cancer Genome Screening System

(SCRUM-Japan) project under the supervision of National Cancer Center Hospital, Japan [136], which assists hospitals and pharmaceutical companies develop PPM for cancer. The aim of SCRUM-Japan trials is to enroll 4750 patients with cancer in about 2 years' plan starting in February 2015 and ending in March 2017 [136].

*E. South Korea*

South Korea introduced itself with the International Precision Medicine Center (IPMC) as the world's first Precision Medicine center focused on Cell Therapy. The IPMC is envisioned to take a pioneering role in standardization of future medicine with a focus on genome and bio convergence technology [137]. The Korean scientists are succeeded to produce a de novo genome assembly for a Korean individuals and the results are published in Nature [138]. The Korea's biobanking system is currently operating a nationwide network of 17 university-affiliated hospitals to collect bio-specimens from patients and then National Biobank of Korea has collected biological samples from some 770, 000 people and distributed them to 1,915 research projects which resulted in a total of 751 research papers as of the end of December 2016 [139], [140]. From industrial sector, the information erupted that the Syapse – a leading precision medicine company joined hands with Seoul National University Hospital (SNUH), Korea to launch a precision oncology program for cancer care improvement in Korea [141]. Moreover, the Korean National Cancer Center with the U.S. National Institutes of Health announced to establish a large-scale precision medicine cohort on cancer [142]. In summary, the Korean government plan is to invest $55.7 million in Precision Medicine until 2021 [143].

*F. Europe*

European Union (EU) is put forwarding numerous efforts to promote precision medicine in the Europe region. As the world's biggest public-private partnership between EU and the European pharmaceutical industry, the Innovative Medicine Initiative (IMI) facilitates collaborations between the stakeholders and provides grants and other financial support to major research projects [144]. IMI in phase 2 that is IMI 2 program (2014-2020), will get a total budget of €3.276 billion, of which €1 billion came from the Health theme of the EU's Seventh Framework Program for Research (FP7) and €1 billion came from in-kind contributions by EFPIA companies [145]. According to a report by ZION, Europe precision medicine market is expected to reach approximately USD 72,800.0 Million by 2022 [146]. Under EU's Horizon2020 Program, Barcelona has started European three-dimensional (3D) genomics project "Multi-scale complex genomics" with a goal is to standardize experiments in 3D genomics and relevant activities like storage of data. The project is allocated a budget of €3 million and will be conducted over three years [147]. The EU funded project "PerMed", where representatives from EU Member States together with other associated countries and stakeholders, have developed a European strategy framework for personalized medicine [6]. PerMed [148] is Coordination and Support Action (CSA) of 27 partners including European key stakeholders and decision makers to allow synergies, avoid duplication, and ensure maximum transparency preparing Europe for leading the global way [149]. The International Consortium for Personalized Medicine (ICPerMed) is a voluntary, EU Member states-led collaboration that brings together over 30 European and international partners to work on coordinating and fostering research to develop and evaluate personalized medicine [150][6].

*G. Australia*

Australia perhaps the world's first country having center specializing in precision medicine for infants and your children which is funded at Murdoch University and have received $473,000 in funding from the WA Department of Health [151]. Precision medicine has the potential to transform Australia's health care system as described in a report released by the Australian Council of Learning Academies (ACOLA) [152]. ACOLA has started a project on precision medicine with a goal to explore the current trends in precision medicine technologies and a broader implementation in the Australian context. In the ACOLA detailed report, there are 12 potential areas are highlighted where precision medicine is likely to show significant impact in the next five to ten years [153]. Australian Genomics is a national network of clinicians, researchers, and diagnostic geneticists and is made up of more than 70 partners organizations with a vision to integrate genomic medicine into healthcare across Australia [154]. National Health and Medical Research Council's (NHMRC) awarded a $25 million grant in 2015 to Australian Genomics for a targeted Call for Research into Preparing Australia for the Genomics Revolution in Healthcare.

Precision Medicine is largely endorsed by other parts of the world such as African, Middle East, and others Asian countries in their own capacity and scope. Orion Health Canada, for instance, has developed a care coordination tool that allows patients to digitally create, update and share their personalized care plan as well as the clinicians are provided with the cognitive support to make the best decisions possible [155][153]. Similarly, the Precision Driven Health initiative (PDHI) in New Zealand is contributing to the growing body of international research to enable the practice of the precision medicine while including genetic data, as well as information from exogenous sources such as an individual's diet and social circumstances [156].

A wide array of international initiatives and consortiums have established to form guidelines for the responsible and homogeneous approach to data movement from one place to another place [5]. For instance, the Global Alliance for

Genomic and Health (GA4GH) is meant to create interoperable technical standards [157].

## VIII. Future Directions

PM is broadly welcomed around the world; the area is however still in its infancy and many aspects are untouched and mount of challenges lie ahead. It is still a big challenge to construct an infrastructure that entirely supports the prevalent sharing and effective use of health and genomic data in order to advance the healthcare system that is least reliant on on external sponsored resources [30].

### A. Challenges and open questions

**Data complexity, volume, and computational challenges.** The computational requirement of molecular and -omics data analysis is huge. The big data analytics is challenging because of multiple factors such as frequency, quality, dimensionality, and heterogeneity [29]. The processing power and memory of personal computers are usually not enough to process DNA sequence data for analysis and interpretations. To support individual researchers for their investigations, need cloud-based computing resources to share the processing power and space. The biomedical data complexity upsurges in dual directions: the number of sample and the heterogeneity [12]. These voluminous complex data are available in different regions of the world through different initiatives using -omic and molecular data capturing technologies which are now becoming faster and cheaper. The variety of available biological data entities for instance genes, proteins, metabolites, drugs, diseases, etc. are so large to manage through basic and simple methods. To handle, process, and annotate such a gigantic and diverse data is not only computationally intensive rather it requires significant computational hardware [158]. For instance, mapping of short reads to get 30x coverage of the human genome, require 13 CPU days. Not only hardware, rather a comprehensive database that contains clinical, genomic and molecular information as much as possible. To deal with the high-throughput data, various method for dimensionality reduction in feature extraction (PCA, SVD, tensor-based approaches [159]) and in feature selection (filter-based and wrapper-based sequential feature selection [160]) are experimented.

Creation of mutation databases challenge. Knowledge bases for example ClinVar and MyCancerGenome are still immature and unfinished thus raises the need to create custom mutation databases by different centers [11]. Also, there is a lack of precise annotations of variants which required databases to contain the curated variants and their interactions with potential drugs [161].

**Integration of heterogeneous data types challenge.** The numerous data types such as omics, molecular, imaging, pathology, physiology, lifestyle, and clinical will be required to incorporated together for predictive models [162]. The orthogonal nature of molecular assays does not allow smooth analysis with clinical data, as a result, separate analysis is performed initially and later they are integrated. This kind of practice is time consuming and it hides the holistic view of data at one place.

**Data privacy challenge.** The protection of genomic data from being used against employment and health protection, various ethical and social issues need to be addressed [163]. It is also required to educate public workforce, develop human capital and infrastructure, and empower the general public with correct information. Moreover, Cloud and Web is likely to play a huge role in the management of massive genomic data and the same time, mobile computing will be used to access those data which increases the privacy concerns [164].

### B. A proposed holistic integrated precision medicine framework

To address the unresolved challenges of PM, more informatics approaches are required to be designed. We designed a futuristic framework (Figure 8), by incorporating functions covering the most needed areas of PM implementation. The framework is a high-level demonstration of modules such as primary analysis, secondary analysis connected with knowledge management and data analytics that produce knowledge and data services respectively. These services are provisioned to use by different stakeholders and organizations including hospitals, pharmacies, and laboratories. The framework has also provision for security and privacy functions to access to the individuals' data through adequate authentication, authorization, and access policy.

The primary analysis module is designed to acquire diverse data from different input sources: clinical data, molecular and -omic data, sensory data, environmental data, and published literature data. At this stage, the data is preprocessed to filter-out the undesirable data items through the application of different natural language preprocessing and other statistical techniques. The primary analysis module utilizes the support of multiple tools particularly, -omic data preprocessing tools such as GMAP, BWA, GATK, and others.

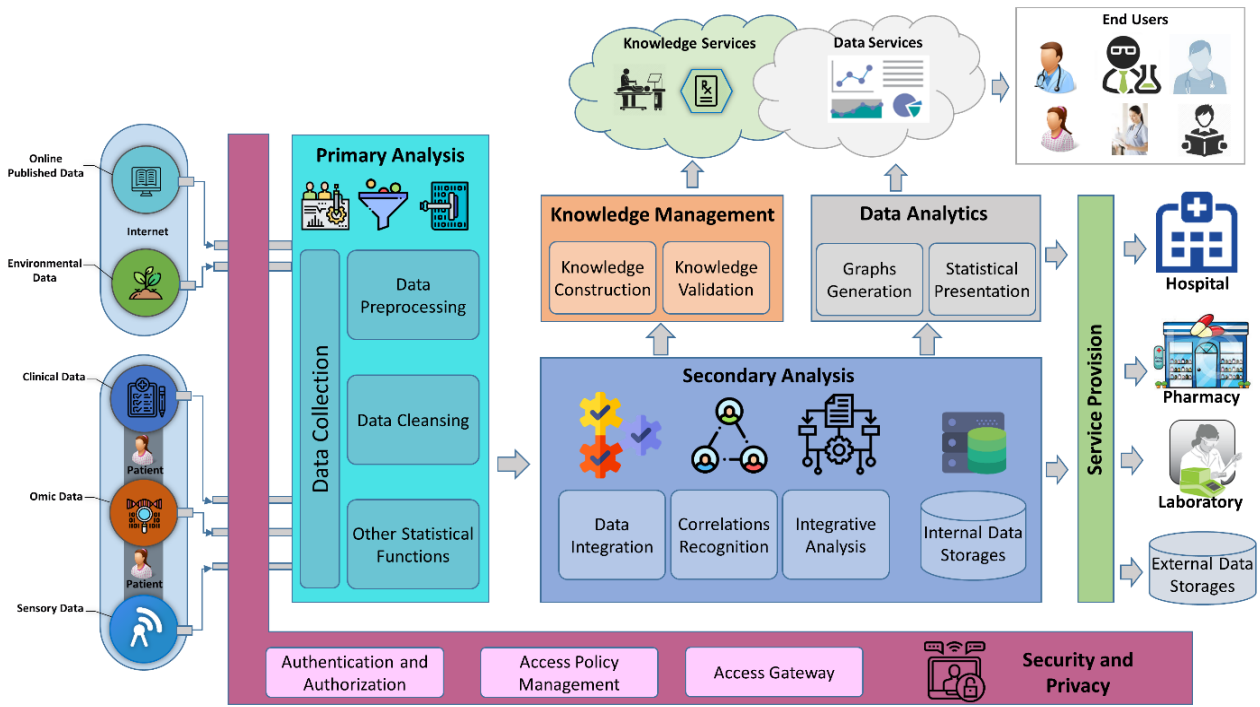

FIGURE 8. *A holistic integrated precision medicine framework*

The secondary analysis module analyzes the data received as an outcome from the primary analysis. Some parts of the data need to be integrated for combine analysis and other may be analyzed independently. One of the important activities is to find correlations among various data items such as among the genes as how they are related while studying a disease occurrence due the genes mutations. Similarly, it takes care of the correlation between genotypes and phenotypes to study the relationships of clinical factors and gene mutations. The analyzed data is stored as an internal storage for further processing as well as it is provided to the external entities as a service.

On the top of secondary analysis, there are two modules: knowledge management and Data Analytics. Both modules utilize the analyzed data generated at the secondary analysis. The knowledge management module constructs KBs by creating, maintaining, and validating knowledge rules from the analyzed data. Based on this knowledge, various knowledge services such clinical decision support services can be produced. Similarly, the data analytics module targets to design models for descriptive, predictive, and prescriptive services. The analytical models generate data visualization services to present data in graphs, charts, and other statistical mode of presentations.

### IX. Conclusions

Both medical professionals and informatics researchers across the globe have started to device computational infrastructural solutions to address the need of timely and precise decision on a patient health issue. It is a high time for both the informatics community and the medical community to collaborate with each other to make a combine effort for achieving the common goal of a better-quality patient care. In this study, we elaborated the major areas of research and development for the realization of PM in the perspective of informatics. The study provides a fair attention to cover the important aspects and requirements to establish the PM program. We explained the need of coexistence of EBM and PM by bridging the gap between them. To understand the informatics viewpoint of how the PM is implemented, we provided an overview of enabling tools and techniques in three potential areas: biomedical informatics, clinical informatics, and participatory health informatics. For a deeper understanding of PM, the paper offers a broad view on how AI and big data become an integral part of PM. We also associated the IoT paradigm with PM and uncovers various advantages of integrating the two approaches. In addition, this paper highlights some of the major implementation challenges in terms of computational tools, data integration, security, standardization, and overall infrastructural solutions that are required to implement PM. Finally, we proposed an integrated holistic framework for PM to overcome the existing limitations. In summary, the outcomes of this study are expected to be beneficial for the researchers and professionals working in the area of medical informatics.

**Conflicts of Interest:** The authors declare no conflict of interest.